# Exact Synthesis of 3-qubit Quantum Circuits from Non-binary Quantum Gates Using Multiple-Valued Logic and Group Theory

Guowu Yang, William N. N. Hung, Xiaoyu Song and Marek Perkowski
Department of ECE, Portland State University, Portland, Oregon, USA
{guowu | whung | song | mperkows} @ ece.pdx.edu

**Abstract**

*We propose an approach to optimally synthesize quantum circuits from non-permutative quantum gates such as Controlled-Square-Root–of-Not (i.e. Controlled-V). Our approach reduces the synthesis problem to multiple-valued optimization and uses group theory. We devise a novel technique that transforms the quantum logic synthesis problem from a multi-valued constrained optimization problem to a group permutation problem. The transformation enables us to utilize group theory to exploit the properties of the synthesis problem. Assuming a cost of one for each two-qubit gate, we found all reversible circuits with quantum costs of 4, 5, 6, etc, and give another algorithm to realize these reversible circuits with quantum gates.*

## 1. Introduction

In this paper, we propose a novel approach to optimally synthesize quantum circuits by group theory where the primary inputs are binary basis states (outputs are not necessarily binary, they may be random binary after measurement of superposition states). Finding the smallest number of gates to synthesize a reversible circuit does not necessarily result in a quantum implementation with the lowest cost (in terms of quantum gates) [2]. The exact minimal costs for NMR [1] realization of several quantum gates from truly quantum (not permutative) gates such as Pauli Rotations or Controlled-Square-Root-of-Not have been calculated [4]. They can be also calculated for other quantum technologies. We focus on synthesizing reversible circuits to quantum implementations with the lowest cost. The method is general and enumerative. It can be adapted to any particular numerical values of costs. These circuits include common reversible gates that can be used at higher levels of logic synthesis or for technology mapping. We formulate the quantum logic synthesis problem via group theory. Our method guarantees to find the minimum quantum-cost implementation with truly quantum gates (given a set of specified component gates). In contrast to previous works, which either use permutative reversible gates to design permutative circuits or universal quantum gates to design quantum circuits, we use a subset of quantum gates to design permutative binary circuits that can be either deterministic (when output symbols are restricted to basis binary states) or probabilistic (when there is no such constraint imposed on the output symbols).

## 3. Formulation

We briefly describe our problem formulation in this section. Further details of our formulation can be found in our technical report [3].

We are interested in synthesizing quantum circuits using elementary quantum gates: NOT gates, XOR (controlled-NOT) gates, controlled-$V$ gates and controlled-$V^+$ gates. In order to use Group Theory, we need to encode the input values. Given our elementary quantum gates, there are four possible values [2] for each qubit: 0, 1, $V_0$, and $V_1$. We represent quantum states as permutations (of truth table entries), and quantum gates as permutations as well. The outputs of quantum gates are simply permutations on permutations. For each gate, we construct a banned set as a filter to extract reasonable truth table entries. A banned set $N_A$, for example, is the set of all entries in which the value of the controlling qubit A is $V_0$ or $V_1$, because only binary values are allowed to be used for control bits.

We constructed a Finding Algorithm (see technical report [3] for details) to compute the reversible circuit set G[k] of all reversible circuits which have cost k. The idea is to create a set A[k] of all quantum circuits that can be constructed using k or less quantum gates. B[k] is the set of quantum circuits that can be constructed using k (and at least k) quantum gates. We create pre_G[k]={b'| b'=Restrictedperm(b,S), b∈B[k]}, where b(S)=S means if the input pattern is binary, then its output is also a binary pattern. So the circuit b' is a reversible circuit with cost ≤ k. We create the set G[k] by subtracting G[k-1], …, G[1] from pre_G[k] because when we compute the

b'=Restrictedperm(b, S) circuit, b' may potentially be a member of any G[j], j<k.

**Theorem 1**. G[k] is the set of all reversible circuits that fulfill our input/output constraints and have minimum quantum cost = k without using NOT gate.

For arbitrary n-qubit reversible circuits, we use N to denote the group realized by NOT gate. The size of N is $2^n$, for all a∈N, a*a=( ), and for all a,b∈N, a*b=( ) iff a=b.

Let G be the set of all n-qubit reversible circuits realized by control gate and Feynman gate. Let H be the set of all n-qubit reversible circuits realized by control gate, Feynman gate and NOT gate. We can deduce that H can be evenly decomposed to $2^n$ leftcosets of G without intersection, shown in the following theorem.

**Theorem 2:**
H= $\cup_{a \in N}$ a*G, and $\forall a,b \in N$ (a≠b) $\Rightarrow$ (a*G)∩(b*G)=∅.

Based on the Finding Algorithm and Theorem 2, we formulate an Expressing Algorithm [3]. An important property of the Expressing Algorithm is that if the quantum cost of g does not exceed the bound cb, then the Expresssing Algorithm will return d[0], d[1], …, d[t] such that g = d[0]*d[1]*…*d[t] and t is minimum.

## 4. Experiments

We applied our minimum cost algorithm to 3 qubit synthesis; the results are shown in the following table.

| Cost k | |G[k]| | $|S_8[k]|$ |
|---|---|---|
| 0 | 1 | 8 |
| 1 | 6 | 48 |
| 2 | 30 | 240 |
| 3 | 52 | 416 |
| 4 | 84 | 672 |
| 5 | 156 | 1248 |
| 6 | 398 | 3184 |
| 7 | 540 | 4320 |

For cost up to 3: G[1], G[2] and G[3] consists of the set of the binary input binary output circuits which are the combinations of 1, 2, 3 Feynman gates respectively. In G[4], there are 60 circuits realized by 4 Feynman gates, the other 24 circuits realized by 3 control gates and 1 Feynman gate. And these 24 circuits exhibit a property of universal gates: all 3-bit binary input and binary output reversible circuits can be realized by NOT gates, Feynman gates and any one of these 24 circuits. There are four representative circuits from these 24 circuits. Each of these four circuits has other five similar circuits with different permutations of the three bits.

Our algorithm can be used to synthesize a quantum circuit from a specified circuit. We applied our algorithm to synthesize the well known Peres and Toffoli circuits. It took 9 CPU seconds (on a 850MHz Pentium® III) to synthesize the Peres circuit (cost=4) and 98 seconds for the Toffoli circuit (cost=5).

Our synthesis algorithm found two implementations for Peres. For the Toffoli circuit, we found four quantum implementations. Notice that for runtime performance our algorithm does not intend to find all possible implementations of the specified circuit. It only finds some implementations as the synthesis result.

Details of our experiments can be found in our technical report [3].

## 6. Conclusion

In this paper we formulated a method to exactly minimize a subset of quantum circuits by reducing the problem to multiple-valued logic and group theory. As far as we know, this is the first time that such a combined approach has been proposed. Using this method we found many new gates and inexpensive realizations of permutative quantum circuits. For instance, we found a family of Peres-like gates which have all the same lowest cost and can be used to synthesize permutative quantum circuits (section 4). Not only is the Peres gate the cheapest of all NMR realized permutative gates, but we show that there is a large family of such gates with the same smallest possible cost, for which nobody has developed a synthesis method yet. Our method can also be used to synthesize circuits with binary inputs and superposed quantum output states. Because such output states are converted in "quantum measurements" to randomly generated vectors with known probabilities [1], our method is then without any modification a new approach to synthesize a class of binary-input circuits that have random but controlled binary outputs (we remove the constraint that outputs are binary states).